\documentstyle[twoside,11pt,aaspp4]{article}
\slugcomment{To appear in {\it The Astronomical Journal}}
\begin{document}

\title{A Search for Photometric Rotation Periods in Low-Mass Stars and
Brown Dwarfs in the Pleiades}

\author{Donald M. Terndrup,
\altaffilmark{1}
\altaffiltext{1}{Department of Astronomy, The Ohio State University,
Columbus, OH 43210; terndrup@astronomy.ohio-state.edu, pinsono@astronomy.ohio-state.edu.}
Anita Krishnamurthi,
\altaffilmark{2}
\altaffiltext{2}{JILA, University of Colorado and National Institute of
Standards and Technology, Boulder, CO 80309-0440;
anitak@casa.colorado.edu.} \\  
Marc H. Pinsonneault,
\altaffilmark{1}
\and
John R. Stauffer
\altaffilmark{3}
\altaffiltext{3}{Smithsonian Astrophysical Observatory, 60 Garden Street,
Cambridge, MA 02138; stauffer@amber.harvard.edu.}
}

\slugcomment{}

\begin{abstract}

We have photometrically monitored (Cousins $I_C$) eight low mass stars
and brown dwarfs which are probable members of the Pleiades.  We
derived rotation periods for two of the stars -- HHJ409 and CFHT-PL8 --
to be 0.258 d and 0.401 d, respectively.  The masses of these stars are
near 0.4 and 0.08 $M_\odot$, respectively; the latter is the second
such object near the hydrogen-burning boundary for which a rotation
period has been measured.  We also observed HHJ409 in $V$; the relative
amplitude in the two bands shows that the spots in that star are about
200 K cooler than the stellar effective temperature of 3560 K and have
a filling factor on the order of 13\%.  With one possible exception,
the remaining stars in the sample do not show photometric variations
larger than the mean error of measurement.  We also examined the M9.5V
disk star 2MASSJ0149, which had previously exhibited a strong flare
event, but did not detect any photometric variation.

\end{abstract}


\section{Introduction}

During the last two decades, there have been many extensive studies of
the rotation rates of G and K stars in open clusters (Stauffer et al.
1984, 1985, 1989; \cite{sh87}; \cite{st91}; \cite{sod93}; \cite{kt98};
\cite{que98}), accompanied by considerable progress in understanding
the mechanisms and mass dependence of angular momentum loss (see
\cite{pin97} for a review).  Briefly, most stars arrive on the main
sequence as relatively slow rotators, but about 20 - 25\% arrive with
rapid rotation ($20 < v \sin i < 200$ km s$^{-1}$).  The number of
rapid rotators and the maximum observed rotation rates decline with
increasing cluster age, showing that the time scale for these rapid
rotators to lose most of their angular momentum ranges from tens to
hundreds of millions of years, and that this time scale becomes longer
with decreasing stellar mass.  In order to be consistent with these
observations, theoretical evolutionary models require one or more of
the following features: (a) core-envelope rotational decoupling
(\cite{end81}; \cite{pin90}); (b) coupling of the stellar rotation to
that of its circumstellar disk during pre-main sequence evolution
(\cite{kon91}; \cite{li93}); and (c) saturation of the angular momentum
loss rate above a specified rotation rate (\cite{kep95}).

Though the general trends are clear at higher masses, there is still
not much information on the rotation rates of main sequence stars of
low  mass (here $M \leq 0.4 M_\odot$).  \cite{jon97}, \cite{sta97}, and
\cite{opp97} have determined rotational velocities ($v \sin i$) for a
small sample of stars in the Pleiades and Hyades with masses down to
$\sim 0.1 M_\odot$, while \cite{bm95} and \cite{del98} have similarly
obtained rotational velocities for low mass field stars.  All of these
studies indicate that the spindown time scales are longer for lower
mass stars, with the time scale near 0.1 $M_\odot$ being several
billion years.  This facet of the observational database is best
explained by assuming that the critical rotation rate for saturation of
the angular momentum loss rate is a function of mass (\cite{kri97};
\cite{bou97}).

Obtaining more rotational data in the low-mass regime would help to
constrain models of angular momentum evolution at the bottom of the
main sequence.  It would also be of interest to extend the samples into
the substellar regime (``brown dwarfs'') to explore whether current
formulations of angular momentum loss for stars are applicable there as
well.  Brown dwarfs are very low mass objects that do not burn
hydrogen, and their properties are not very well characterized.
Determining rotation rates for low-mass stars and substellar objects
using high-resolution spectra is extremely challenging as they are
quite faint ($I > 17$) and their spectra (late M-type) are complex.
Furthermore, stars at and below the hydrogen-burning mass limit have
quite small radii; the smallest rotation rates that are routinely
determined spectroscopically (say $\approx 5$ km s$^{-1}$), correspond
to short rotation periods, about 0.5 day.  Therefore the discovery
of low-mass stars and brown dwarfs with periods in excess of 0.5 -- 1
day may best be accomplished through through photometry, which can
detect brightness modulation from starspots.  A determination
of rotation periods by this method for brown dwarfs would indicate the
presence of spots and hence magnetic activity on these sub-stellar
objects, since the presence of starspots is a manifestation of magnetic
fields.

Motivated by these considerations, we undertook a pilot study to derive
rotation periods from photometric modulation of low-mass members of the
Pleiades.  Section 2 describes the survey, and data analysis
techniques, while $\S~3$ presents our results and a discussion. 

\section{Observations and Analysis}

We selected eight low-mass stars and brown dwarf candidates in the
Pleiades; a list of these stars and a compilation of some of their
properties is in Table 1.  Columns 2--3 of that table show the J2000.0
positions for the targets, followed in columns 4--7, respectively, by
their spectral classification and available photometry.  The last two
columns give a mass estimate, as derived from comparison to theoretical
isochrones, and references for the various data about each star.  The
full sample ranges from $M \approx 0.4M_\odot$ down to $M \approx 0.05
M_\odot$.

The two Pleiades stars with $M \approx 0.4 M_\odot$, HHJ 409 and HHJ
375 (\cite{hhj}), were selected for monitoring because they are known
to have quite different spectroscopic rotation rates of $v \sin i$ = 69
km s$^{-1}$ and 10 km s$^{-1}$, respectively (\cite{sta99}).  The rest
of the stars were selected from deep imaging surveys which have
produced candidate Pleiades brown dwarfs:  Teide 1 (\cite{rebolo95}),
Teide 2 (\cite{martin98}), Roque 11 and Roque 13 (\cite{zap97}), and
CFHT-PL-8 and CFHT-PL-12 (\cite{bouv98}).

A number of the fainter stars have spectra which show measurable
equivalent widths of \ion{Li}{1}\ $\lambda 6708$ \AA.  As originally
proposed by \cite{mag93}, stars with masses below about 0.065 $M_\odot$
never develop central core temperatures sufficient to destroy lithium
($3 \times 10^6$ K).  At higher masses, the time it takes the core to
achieve this temperature is a sensitive function of mass.  Because
low-mass stars are fully convective, the surface Li is rapidly depleted
once the core reaches a high enough temperature.  For clusters of the
age of the Pleiades (100 -- 120 Myr), the mass above which Li is
depleted is about 0.08 $M_\odot$ (\cite{sta98} and references
therein);  consequently, finding a Li abundance which does not show
significant depletion would show that a given star 
in the Pleiades is a brown dwarf.
The stars with spectroscopic confirmation of their status as brown
dwarfs are Teide 1 (\cite{rebolo96}), Teide 2 (\cite{martin98}), Roque
13 and CFHT-Pl 12 (\cite{sta98}).

We also monitored the very low-mass object 2MASS J0149090+295613
(hereafter referred to as 2MASSJ0149), which has exhibited a
significant flare in H$\alpha$ (\cite{liebert99}).  As this flare can
be interpreted as evidence for magnetic activity, we included this
object in our sample to determine if we could observe spot modulation,
which could be another indicator (or confirmation) of magnetic
activity.

We used the 2.4-m Hiltner telescope of the MDM Observatory on Kitt Peak
to obtain $I_C$ (Cousins $I$ band) photometry of our targets from UT
11-23 November 1998.  The detector was the ``Wilbur'' CCD, a thick
$2048 \times 2048$ Loral chip which was operated at a nominal gain of
2.3 electrons/DN and a readout noise of 4.7 electrons r.m.s.  The
readout was binned by a factor of two in rows and columns to yield a
pixel scale of 0.393 arcseconds/pixel.  In addition, we obtained a few
observations of HHJ~409 in $V$ to measure the relative photometric
amplitude at two wavelengths.  Most of the run was photometric, with
median seeing about $0.8\arcsec$; on a few nights we observed a large
number of standards from \cite{landolt} for calibration, though in this
paper we will only discuss amplitude variations and will not present
calibrated photometry.

The EXPORT version of Sun/IRAF V2.11.1 for SunOS 4 and Solaris 2.6 was
used for the initial processing of data.  The raw images were corrected
for overscan bias, zero-exposure level, and were flattened by
twilight-sky flats.  Inspection of the flats from several nights and of
the sky level on deep exposures showed that the flattening process was
accurate to 0.2\% over most of the frames, increasing to about 1\% at
the extreme edges of the frames.

The observing strategy was to make repeated {\it visits} to each target
and obtain from 2-5 exposures in quick succession; exposures ranged
from 5-10 sec for the bright stars HHJ~375 and HHJ~409, and were as
long as 300 sec for the faintest stars in the sample.  Because the
target fields were uncrowded, aperture photometry (rather than
PSF-fitting) was adequate.  Magnitudes were obtained for each
unsaturated star on each frame in an aperture of radius 3 times the
stellar FWHM; the median sky was measured in an annulus with radii of 5
and 7 times the FWHM.

The assembly of the magnitudes into a time series was accomplished in
two steps.  In the first, the magnitudes from the repeated exposures at
each visit were averaged after rescaling by an additive constant.  This
offset was determined by shifting the stellar coordinates to a common
system, then determining the mean weighted magnitude difference between
photometry from pairs of frames, using all unsaturated stars common to
each frame.  The weights were assigned as $1/\sigma^2$, where $\sigma$
was the magnitude error reported from the aperture photometry
software.  Because there were always a number of well-exposed stars on
each frame, the magnitude offset could be determined very precisely
(error typically $\leq 0.004$ mag).  The target star was included in
the determination of the magnitude offset at this step because it would
not have varied in the $\leq 30$ min spent at each visit.  The average
HJD was taken as the time point for each visit.

In the second step, the average photometry for the many visits to each
star were similarly scaled using mean weighted magnitude differences,
but this time the target star was not included in the determination of
the offset.  Errors in the average magnitude were computed in both
steps by taking the larger of the actual weighted error in the mean, or
the expected error for Gaussian statistics.

Periods were determined using the \cite{sca82} implementation of the
{\small PERIODOGRAM} algorithm for unevenly spaced data, which also
yields the ``false alarm probability'' (FAP) that random fluctuations
would produce the derived period.  We also confirmed the derived
periods (or absence of a detected signal) using the algorithm described
in \cite{rld87} (kindly supplied by D.\ H.\ Roberts), which takes into
account aliasing introduced by the sampling frequency.  Both algorithms
yielded the same set of stars with definite detections and the
resulting periods were the same.

\section{Results and Discussion} 

The results of our observational program are summarized in Table 2.
The first column lists the star names as in Table 1, while columns 2--5
show, respectively, the number of visits to the star, the r.m.s.
scatter about the mean magnitude, and the average error in a single
measurement.  The last two columns display the derived period in days
for the star and the false-alarm probability for the stars with
definite periods.

As discussed in Section 2, the targets in our sample were selected to
have a mass distribution which crossed the H-fusion boundary.  We were
able to measure definite rotation periods for two of the four low-mass
stars (hydrogen burning) in our sample, namely HHJ 409 and CFHT-PL-8.
The photometry for these two stars, phased to the derived period, is
shown in Figures 1 and 2.  Also shown on each figure is the photometry
for a nearby star of similar brightness phased to the period of the
target star;  the scatter in the photometry of the nearby star is equal
to the expected photometric error, verifying that the brightness
variations in the target are real.  In both figures, the differential
photometry is plotted in the sense ($\Delta I_C = I_C - \langle
I_C\rangle$). There is insufficient time-series data to be able to find
a period for HHJ 375.  This star shows photometric variation at the
2.5$\sigma$ level, and a very approximate period of $\sim$1.5d (FAP
$\sim 0.4$).  CHFT-PL-8 has an amplitude of 10\%;  this star is
comparable in mass to a star in the young cluster Alpha Perseii for
which \cite{mzo97} derived a photometric rotation period. None of the
other stars showed variations which were as much as twice the typical
error of measurement (see table 2).  Thus, we were unable to measure
rotation periods for any of the objects which are likely to be
sub-stellar.

We also followed HHJ 409 in $V$ (five visits) to determine the relative
amplitude compared to that in $I_C$.  Figure 3 plots the difference in
magnitude with respect to the average magnitude in each filter for
these five visits.  We find that the relative amplitude in $V$ is 2.1
times that in $I_C$, and since the the star gets redder as it gets
fainter, the spots are cooler than the average temperature of the
photosphere.  The $V-I$ color of the star indicates a photospheric
temperature of $T = 3560$ K.  To estimate roughly the temperature
difference between the spots and the star, we did a simple model of
blackbody spots (see figure 4).  For each assumed spot temperature, we
derived a filling factor which would produce an amplitude of 0.04 mag
in $I_C$; the cooler the spots, the smaller the required filling factor
to produce this amplitude.  Then for each temperature, we calculated
the resulting expected $V$ amplitude.  At spot temperatures near 3330 K
and filling factors near 13\%, the observed $V$ amplitude of 0.08 mag
was matched (Figure 4).

Though this measurement of the relative amplitudes in the two filters
was only done for one star, it suggests that observations in
wavelengths as red as $I_C$ may be quite useful for this kind of work,
for the simple reason that the stars are so much brighter in $I$ than
in $V$ that it is easier to obtain good signal-to-noise in the
photometry.  Still, only two of our stars had amplitudes which were
larger than the errors in the photometry.  Several alternative
explanations arise:  the non-variables may have very little
chromospheric activity, which is strongly correlated with rotation
(e.g., \cite{kt98} and references therein). The stars may have activity
cycles and it so happened that they did not have spots with a large
filling factor during the time of our observations.  Some higher-mass
Pleiades stars show significant variations in their amplitude, which is
interpreted as being the result of activity cycles (e.g., \cite{sta87b};
\cite{pro93}), and our target J0149 has been quiescent except for the
one flare event (\cite{liebert99}).  Finally, the spots may have very
low contrast to the average photospheric color.

There have been several efforts recently to measure activity in brown
dwarfs.  \cite{neu99} conducted a search for X-ray emission from
confirmed (Li test) and candidate brown dwarfs.  A very small number of
the targets were detected in X-rays: only one out of 26 brown dwarfs
was detected, while 4 out of 57 of the brown dwarf candidates were
detected.  Except for 2MASSJ0149, the brown dwarfs and brown dwarf
candidates studied in this paper were all included in the \cite{neu99}
study.  No detections were made and only upper limits were obtained for
those of our targets included in that study.

This lack of activity is consistent with the findings of \cite{kri99},
who conducted a deep search for radio emission from a sample of very
low mass stars and brown dwarfs in the field; no radio emission was
detected from the studied targets.  A possible explanation for a lack
of detectable activity in such low-mass objects has been suggested by
\cite{neu99} - brown dwarfs are fully convective objects that briefly
burn deuterium and are not massive enough to burn hydrogen.  When these
objects stop burning deuterium, the core starts to cool down. As the
temperature difference between the core and the surface decreases, the
convective velocities decrease.  This dampens the dynamo, which is
powered not only by rotation, but also by the convection velocity.
Thus, the objects may be rotating rapidly (as is seen from some
measurements) but still not show much sign of activity in the form of
X-ray emission, radio emission or spots.

In summary, we monitored a set of low-mass stars and brown dwarfs in
the Pleiades and one low-mass star in the field in order to determine
photometric rotational periods.  We were able to derive rotation
periods for two of the low mass stars.  We thus verify (as in
Mart\'{\i}n \& Zapatero Osorio 1997) that at least some stars near the
hydrogen burning limit have large spot groups.  The frequency of the
spots is low (about 20-25\%), or the spots have temperatures which are
near that of the photosphere in these cool stars.  We were unable to
measure rotation periods for any of the (likely) substellar objects.
From comparison to earlier work looking for activity, we speculate that
this lack of detectable activity may be an inherent problem with using
a sample of objects that are older than a few million years.  If this
hypothesis is correct, only very young brown dwarfs show signs of
activity and it will be necessary to study such young objects to be
able to detect spot activity.  We will identify and study such
candidates in an effort to determine if brown dwarfs show signs of
magnetic activity at younger ages.

\acknowledgments  

We acknowledge support from NSF grant AST-9731621 to MHP and DT.
Thanks to MDM staff, and to Liam Mac Iomhair for his assistance at the
telescope.  AK received support from NASA grant S-56500-D to the
University of Colorado.

%******************************tables*****************************************

% Terndrup et al., Table 1
\begin{deluxetable}{lccccccl}
\tablenum{1}
\tablewidth{0pt}
\tablecaption{Characteristics of observed targets}
\tablehead{
\colhead{Name} & 
\colhead{RA (2000)} & 
\colhead{Dec (2000)} & 
\colhead{Spectral type}& 
\colhead{$I$} & 
\colhead{$I-K$} &
\colhead{Mass ($M_\odot$)\tablenotemark{a}} &
\colhead{Reference\tablenotemark{b}}
}
\startdata

2MASSJ0149  & 01:49:09.0 & +29:56:13 & M9.5V  &\nodata& 1.40    & \nodata &  L \nl
Roque 11    & 03:47:12.1 & +24:28:32 & M8     & 18.75 & 3.63    &  0.06   &  Z3\nl
Roque 13    & 03:45:50.6 & +24:09:03 & M7.5   & 18.25 & 3.65    &  0.07   &  Z3\nl
Teide 1	    & 03:47:18.0 & +24:22:31 & BD     & 18.80 & 3.69    &  0.055  &  Z1\nl
Teide 2	    & 03:52:06.7 & +24:16:01 & M6     & 17.82 & \nodata &  0.072  &  M \nl
HHJ 375     & 03:48:05.8 & +23:02:04 & \nodata & 13.74 & \nodata &  0.39   &  H,S \nl
HHJ 409     & 03:42:40.2 & +23:59:22 & \nodata & 13.34 & \nodata &  0.39   &  H,S \nl
CFHT-PL8    & 03:42:26.8 & +24:50:21 & \nodata & 17.42 & \nodata &  0.084  &  B  \nl
CFHT-PL12   & 03:53:55.1 & +23:16:01 & \nodata & 18.02 & \nodata &  0.070  &  B \nl

\enddata
\tablenotetext{a}{Masses estimated from photometry; see references.}
\tablenotetext{b}{References: (B) Bouvier et al. (1998); Hambly et al.
(1993); (L) Liebert et al. (1999); 
(M) Mart\'{\i}n et al. (1998); (S) Stauffer et al. 1999; 
(Z1) Zapatero Osorio et al. (1997); 
(Z3) Zapatero Osorio et al. (1999).}
\end{deluxetable}

% Terndrup et al., Table 2
\begin{deluxetable}{lccccc}
\tablenum{2}
\tablewidth{0pt}
\tablecaption{Results}
\tablehead{
\colhead{ } & 
\colhead{ } & 
\colhead{r.m.s}& 
\colhead{mean} & 
\colhead{period} &
\colhead{ } \\
\colhead{Name} & 
\colhead{Visits} & 
\colhead{$\Delta I_C$}& 
\colhead{error} & 
\colhead{(day)} &
\colhead{FAP}
}
\startdata

2MASSJ0149  & 37  & 0.014 & 0.011 & \nodata & \nodata  \nl
Roque 11    & 27  & 0.041 & 0.043 & \nodata & \nodata  \nl
Roque 13    & 21  & 0.023 & 0.025 & \nodata & \nodata  \nl
Teide 1	    & 25  & 0.045 & 0.027 & \nodata & \nodata  \nl
Teide 2	    & 31  & 0.018 & 0.014 & \nodata & \nodata  \nl
HHJ 375     & 18  & 0.007 & 0.004 & \nodata & \nodata  \nl
HHJ 409     & 32  & 0.015 & 0.004 & 0.258   & $3.5 \times 10^{-4}$  \nl
CFHT-PL8    & 34  & 0.028 & 0.007 & 0.401   & $5.7 \times 10^{-4}$  \nl
CFHT-PL12   & 32  & 0.016 & 0.012 & \nodata & \nodata  \nl

\enddata
\end{deluxetable}

%******************************figures*****************************************

\clearpage

\begin{figure}  
\plotone{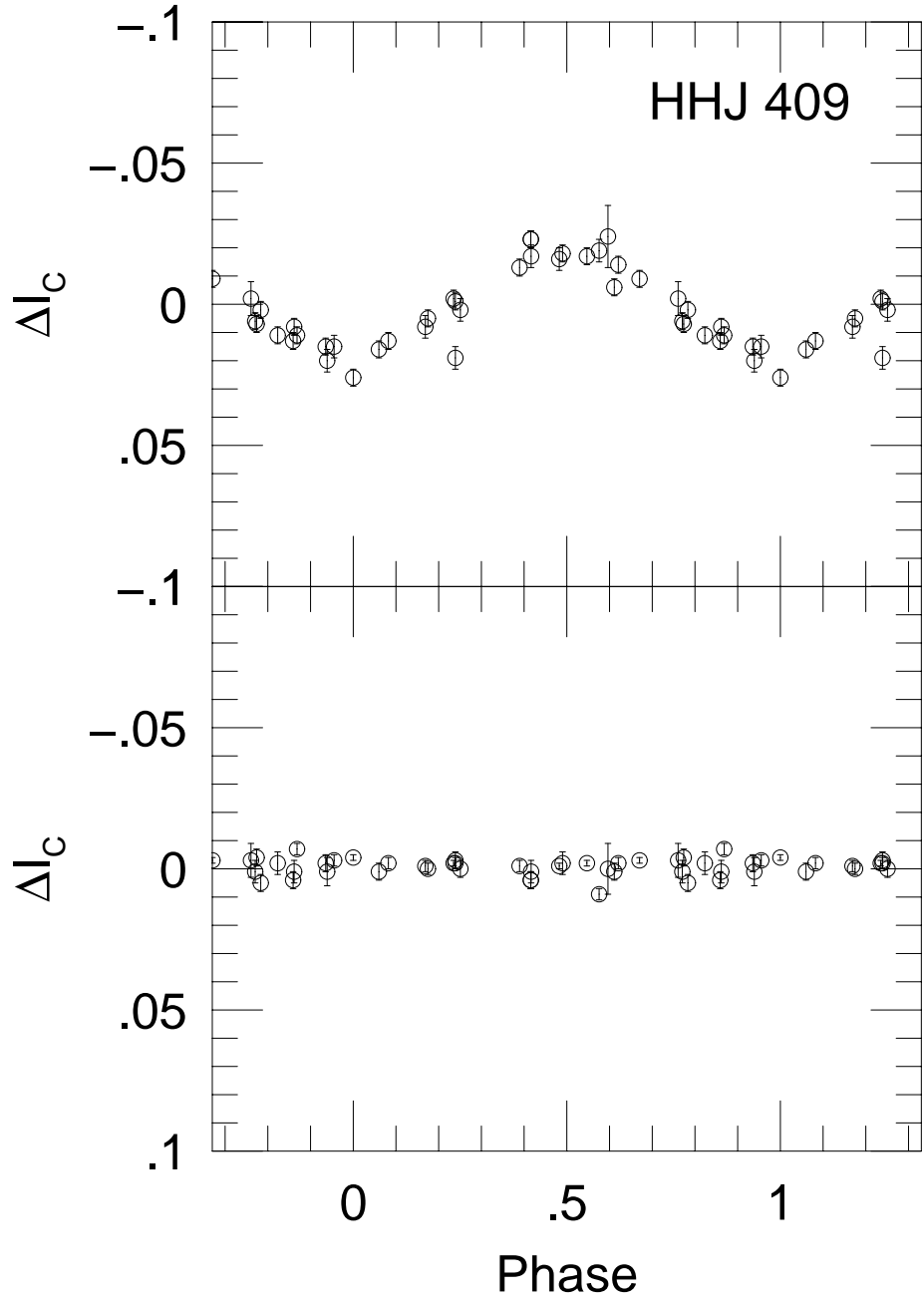}
\caption{Photometry for HHJ 409 (upper panel) and for a
nearby star of similar brightness (lower panel), both phased
to the derived period of 0.258 d.}
\end{figure} 

\begin{figure}  
\plotone{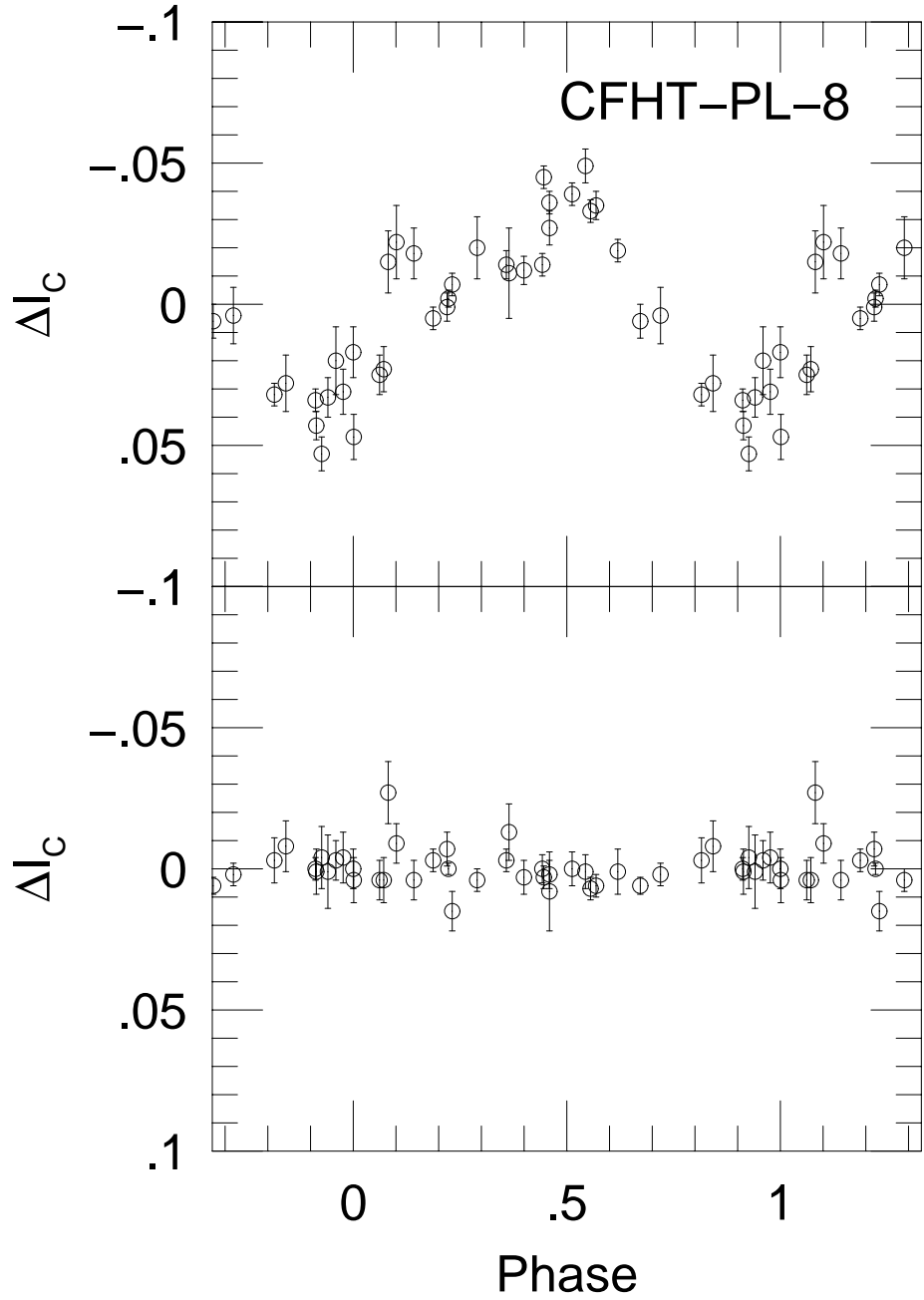}
\caption{Same as Figure 1, but for CFHT-PL-8.  The phasing
is for the derived period of P = 0.402 d.} 
\end{figure} 
 
\begin{figure}  
\plotone{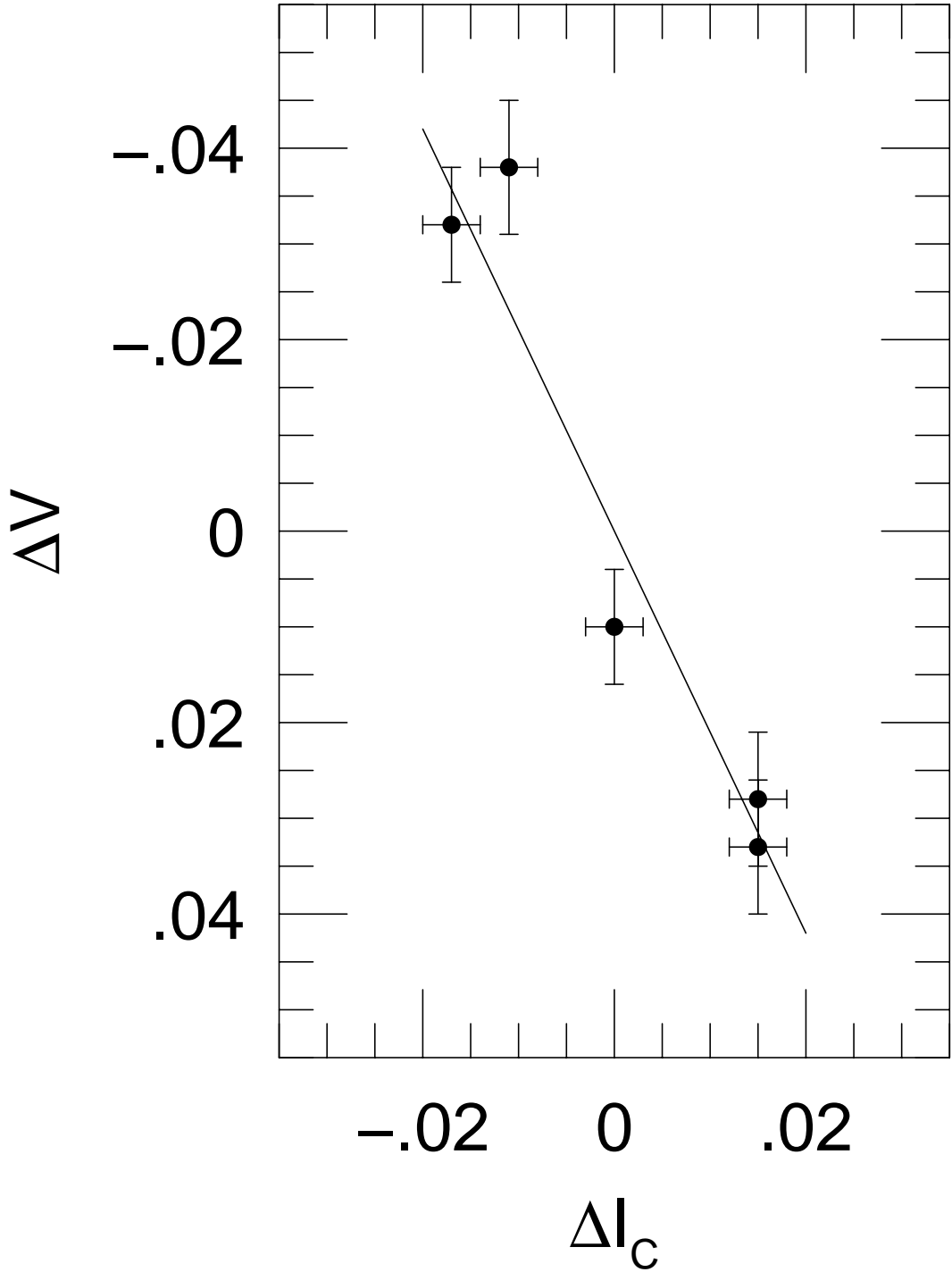}
\caption{Differential photometry in $V$ and $I_C$
for HHJ 409, where the values are in the sense magnitude - mean.
The solid line is a linear fit to the data;  the amplitude in
$V$ is 2.1 times that in $I_C$.  The star becomes redder as it
gets fainter, showing that the spots are cooler than the photosphere.}
\end{figure} 

\begin{figure}  
\plotone{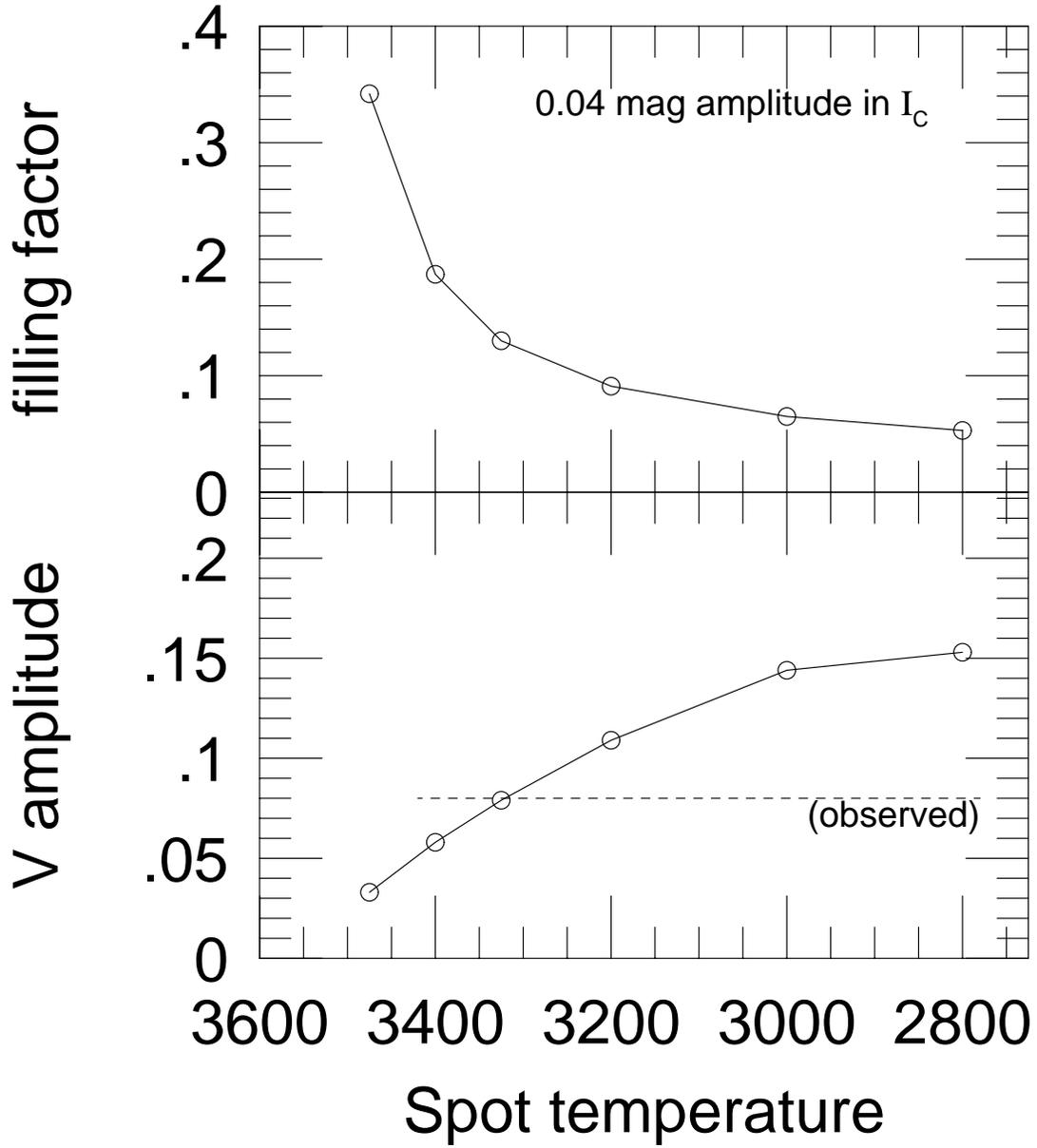}
\caption{Simple model of blackbody spots for HHJ 409: top panel shows
the filling factor required to produce a 4\% variation in $I_{c}$ for
different spot temperatures.  The bottom panel shows the expected $V$
amplitude for different spot temperatures.  The spot temperature for HHJ
409 is taken to be that which corresponds to the observed $V$ amplitude of
0.08 magnitude, or 3330K.  This also corresponds to a filling factor
of $\sim$13\% from the top panel.}
\end{figure} 

\end{document}